\documentclass[a4,12pt,eclepsf]{article}
\usepackage[pdftex]{graphicx} 
\usepackage{wrapfig, fancybox, fancyhdr, amsmath, amssymb, makeidx, 
comment, multirow, bm, here}

\newcommand{\BM}{\begin{minipage}}
\newcommand{\EM}{\end{minipage}}

\newcommand{\DS}{\displaystyle}

\begin{document}

\renewcommand{\theequation}
{\thesection.\arabic{equation}}
\thispagestyle{empty}
\vspace*{5mm} 

\begin{center}
{\LARGE {\bf Universal terms for holographic}} \\[5mm]
{\LARGE {\bf entanglement entropy in}} \\[5mm]
{\LARGE {\bf noncommutative Yang--Mills theory}} \\[5mm]

\vspace*{25mm}

{\Large Tadahito NAKAJIMA} \footnotemark[1]
\vspace*{15mm}

$\footnotemark[1]$ {\it College of Engineering, Nihon University, 
Fukushima 963-8642, Japan} \\[25mm]

{\bf Abstract} \\[10mm]

\end{center}

In this paper, we derive the universal (cut-off-independent) part of the holographic entanglement entropy in the noncommutative Yang-Mills theory and examine its properties in detail. The behavior of the holographic entanglement entropy as a function of a scale of the system changes drastically between large noncommutativity and small noncommutativity. The strong subadditivity inequality for the entanglement entropies in the noncommutative Yang-Mills theory is modified in large noncommutativity. The behavior of entropic $c$-function defined by means of the entanglement entropy also changes drastically between large noncommutativity and small noncommutativity. In addition, there is a transition for the entanglement entropy in the noncommutative Yang-Mills theory at finite temperature.

\clearpage

\setcounter{section}{0}
\section{Introduction}
\setcounter{page}{1}
\setcounter{equation}{0}

The noncommutative gauge theories discussed in this paper is a gauge theories in which the product of any two fields is given by the Moyal-Weyl product \cite{MRS, AA}
\begin{align}
\label{101}
f \star g(x) \equiv f(x) \exp \left(\dfrac{i}{2}\theta^{\mu\nu}
\overleftarrow{\partial_{\mu}} \, \overrightarrow{\partial_{\nu}} \right) g(x)\;,
\end{align}
where $\theta^{\mu\nu}$ is the deformation parameter. It has well known that these gauge theories naturally arise as low energy theories of D-branes in Neveu-Schwarz-Neveu-Schwarz (NS-NS) $B$-field background \cite{CDS, DH, AASJ, SW}. A remarkable phenomenon in these gauge theories is UV/IR mixing \cite{MRS, AA}, where the ultraviolet (UV) and infrared (IR) degrees of freedom of the theory are mixed in a complicated way. It is very important to deepen the understanding such quantum effects not only in perturbative approach but also in nonperturbative approach.

There exists a holographic description for the strongly coupled noncommutative gauge theories in the large $N$ limit \cite{HI, MR, AOSJ, MLYSW}. The holographic description of the noncommutative gauge theories is often useful to investigate how the noncommutativity (the deformation parameter) affect the quantum properties of the gauge theories. For instance, the noncommutativity modifies the Wilson loop behavior \cite{dhar_kita, lee_sin, taka_naka-suzu} and glueball mass spectra \cite{NST}. The holographic duals of noncommutative gauge theories with flavor degrees of freedom have also been constructed by using probe techniques \cite{APR}. Employing the holographic description, we have been able to find the noncommutativity is also reflected in the flavor dynamics \cite{TN_YO_KS}. It should be emphasized that the noncommutativity can also modify phase diagram as for instance chiral symmetry breaking in the noncommutative gauge theory \cite{TN_YO_KS2, TN_YO_KS3}.

In this paper, we focus on quantum entanglement in noncommutative gauge theory.
Entanglement entropy is known as a measure for entanglement in quantum systems (see e.g. \cite{PCJC}). The entanglement entropy of a subsystem $A$ is defined by the von Neumann entropy of the reduced density matrix $\rho_{A}$ of the system $A$, 
\begin{align}
\label{102}
S_{A}=-{\rm tr}_{A}\left(\rho_{A} \ln \rho_{A} \right) \;.
\end{align}
It is possible to compute the entanglement entropy by employing the holographic approach. Ryu and Takayanagi conjectured the holographic formula of entanglement entropy should be 
\begin{align}
\label{103}
S_{A}=\dfrac{{\cal A}}{4G} \;,
\end{align}
where ${\cal A}$ is the area of a minimal surface with a given boundary \cite{SRTT1, SRTT2}. The proof of this formula is given by \cite{LAJM}. 

Quantum physics allows a superposition of states, causing a nonlocal correlation between subsystems far apart from each other. Entanglement is the distinctive concept of the quantum physics, including quantum field theories, and that is one of the important concepts to understand quantum aspects of the quantum physics.

It is known that entanglement entropy for nonlocal field theories whose action contains infinite derivatives follows a volume law, in general \cite{NSTT, UKCNDSJSMW, DWP}. The noncommutative gauge theory is a kind of nonlocal field theories. It would be worth investigating how nonlocality of the noncommutative gauge theory affects the properties of the entanglement entropy. The entanglement entropies in the noncommutative gauge theories has been studied based on the holographic approaches \cite{LBCF, WFAKSK, JLKCR, TJZX}. It have been pointed out that the divergence (cut-off dependence) part of the entanglement entropy in the large noncommutativity limit follows a volume law \cite{LBCF}.

The holographic entanglement entropies for quantum field theories are often regularized for finite by introducing a cut-off parameter. Little attention, however, has been given to the cut-off independent part of the holographic entanglement entropies in the noncommutative gauge theories. In this paper, we try to derive the universal (cut-off-independent) part of the holographic entanglement entropy in the noncommutative Yang-Mills theory and discuss its properties. The properties of the holographic entanglement entropy in the noncommutative Yang-Mills theory should be discussed on the basis of universal (cut-off-independent) quantities.

The paper is organized as follows. In section 2, we introduce the holographic entanglement entropy conjectured by Ryu and Takayanagi and derive the universal (cut-off-independent) part of the holographic entanglement entropy in the noncommutative Yang-Mills theory. In section 3, we investigate the strong subadditivity for the holographic entanglement entropy in the noncommutative Yang-Mills theory. The properties of the mutual information written by the entanglement entropies is also discussed. In section 4, we investigate the properties of the entropic $c$-function in the noncommutative Yang-Mills theory. In section 5, we derive the universal part of the holographic entanglement entropy in the noncommutative Yang-Mills theory at finite temperature and discuss a kind of the transition based on the holographic entanglement entropy. Section 6 is devoted to conclusions and discussions.

%
%

\section{Holographic entanglement entropy in noncommutative Yang-Mills theory} 

\setcounter{equation}{0}
\addtocounter{enumi}{1}

We consider the dual description of the noncommutative Yang-Mills theory on  a spacetime $\mathbb{R}^{1+1} \times \mathbb{R}_{\theta}^{2}$. where $\mathbb{R}_{\theta}^{2}$ is the noncommutative (Moyal) plane defined by a Moyal algebra $[x_{2},\;x_{3}]=i\theta$.  At Large $N$ and strong 't Hooft coupling, a holographic description of the noncommutative Yang-Mills theory is given by
\begin{align}
\label{201}
ds^{2} &=R^{2}
\Bigl[u^{2} \bigl\{dx_{0}{}^{2}+dx_{1}{}^{2} + h(u)(dx_{2}{}^{2}
+dx_{3}{}^{2}) \bigr\} 
+ \left(\dfrac{du^{2}}{u^{2}}+d\Omega_{5}{}^{2} \right) \Bigr] 
\,, \nonumber \\[-3mm]
\quad \\[-3mm]
h& =\dfrac{1}{1+a^{4}u^{4}}\;,
\nonumber 
\end{align}
where $R^{4}=4\pi g_{s} N l_{s}^{4}$ and $a$ denotes the noncommutativity parameter with dimension of length.

We will use the generalized Ryu-Takayanagi formula for the ten-dimensional geometry with a varying dilaton. The holographic definition of entanglement entropy is given by
\begin{align}
\label{202}
S_{A} = \dfrac{{\cal A}}{4G_{N}^{(10)}}
 = \dfrac{1}{4G_{N}^{(10)}}\int d^{8}\sigma e^{-2\Phi}\sqrt{G_{\rm ind}^{(8)}}\;,
\end{align}
where $G^{(10)}_{N}=8\pi^{6}\alpha'^{4}$ is the ten dimensional Newton's constant. The 5-dimensional Newton's constant $G^{(5)}_{N}$ is proportional to  $G^{(10)}_{N}$ up to a volume factor $G^{(5)}_{N}=G^{(10)}_{N}/\pi^{3}R^{5}$. 

Let us compute the entanglement entropy for an infinite strip specified by 
\begin{align}
\label{203}
y \equiv x_{2} \in \left[-\dfrac{l}{2},\;\dfrac{l}{2}\right]\,,
\qquad 
x_{1},\;x_{3}  \in \left(-L, \; L \right) \;.
\end{align}
with $L \to \infty$. It would be worth pointing out that when we exchange $x_ {2}$ for $x_ {3}$, then the noncommutative deformation has no effect on the entanglement entropy. Under this configuration, the entanglement entropy defined by (\ref{202}) takes the form, 
\begin{align}
\label{204}
S_{A}&= \dfrac{N^{2}L^{2}}{\pi} \int du \,
u^{3}\sqrt{y'(u)^{2}+\dfrac{1}{u^{4}h(u)}} \,
\end{align}
where $y'(u)$ is the derivative of $y$ with respect to $u$. We find the quantity $\dfrac{u^{3}y'(u)}{\sqrt{y'^{2}(u)+1/u^{4}h(u)}}$ is a constant which does not depend on $u$. This quantity leads to 
\begin{align}
\label{205}
y'(u)=\dfrac{1}{u^{2}\sqrt{h(u)}}
\dfrac{1}{\sqrt{\dfrac{u^{6}}{u_{\ast}^{6}}-1}} \;,
\end{align}
where $u_{\ast}$ denotes an integral of motion and $u=u_{\ast}$ represents the point of closest approach of the extremal surface. Such surfaces have two branches, joined smoothly at $u=u_{\ast}$ and $u_{\ast}$ can be determined using the boundary conditions:
\begin{align}
\label{206}
y(u_{\ast})=\pm \dfrac{l}{2} \;.
\end{align}
The entanglement entropy given by (\ref{204}) at the stable solution is given by
\begin{align}
\label{207}
S_{A} = \dfrac{N^{2}L^{2}}{\pi}\int^{u_{\Lambda}}_{u_{\ast}} 
du\;u^{4} \sqrt{\dfrac{1+(au)^{4}}{u^{6}-u_{\ast}^{6}}} \;,
\end{align}
where $u_{\Lambda}$ is a cutoff parameter. The (dimensionless) entanglement entropy functional defined by ${\cal S}_{A} \equiv (\pi a^{2}/N^{2}L^{2})\,S_{A}$ can be rewritten as 
\begin{align}
\label{208}
{\cal S}_{A} 
= (au_{\ast})^{2}\int^{1}_{u_{\ast}/u_{\Lambda}} \dfrac{dt}{t^{5}}
\sqrt{\dfrac{t^{4}+(au_{\ast})^{4}}{1-t^{6}}} \,
\end{align}
where $t \equiv u_{\ast}/u$ is a dimensionless variable. The ratio of the length $l$ to the noncommutativity parameter $a$ is also a function of the product of the noncommutativity parameter $a$ and the integral of motion $u_{\ast}$:
\begin{align}
\label{209}
\dfrac{l}{a} = \dfrac{2}{a}\int^{\infty}_{u_{\ast}}du\, y'(u)
= \dfrac{2}{au_{\ast}}
\int^{1}_{0}dt\; t\,\sqrt{\dfrac{t^{4}+(au_{\ast})^{4}}{1-t^{6}}}\;.
\end{align}

In the deep infrared bound $au_{\ast} \cong 0$, we can approximate the right hand side of (\ref{209}) by $\DS{\dfrac{2}{au_{\ast}} \int^{1}_{0}dt\; t\,
\sqrt{\dfrac{t^{4}}{1-t^{6}}}}$, and we have
\begin{align}
\label{210}
\dfrac{l}{a} \cong 2\sqrt{\pi}\,\dfrac{\Gamma(\frac{2}{3})}{\Gamma(\frac{1}{6})} \cdot \dfrac{1}{au_{\ast}} \;,
\end{align}
where $\Gamma$ denotes the Gamma function. The length $l$ given by (\ref{210}) is the same as that in the commutative ($a=0$) version. Hereafter, we refer to the approximation in the deep infrared bound as the {\it commutative regime}. The commutative theory ($a=0$) and the noncommutative ($a \neq 0$) theory can be compared through the approximation in the deep infrared bound.

The entanglement entropy functional (\ref{208}) in the commutative regime can be divided into a universal(finite) part ${\cal S}_{U}$ that is independent of the cutoff parameter $u_{\Lambda}$ and the divergence part ${\cal S}_{D}$ that depends on the cutoff parameter $u_{\Lambda}$:
\begin{align}
\label{211}
{\cal S}_{U} \left(= \dfrac{\pi a^{2}}{N^{2}L^{2}}\,S_{U} \right) 
\cong -\dfrac{\sqrt{\pi}}{2}\dfrac{\Gamma(\frac{2}{3})}{\Gamma(\frac{1}{6})} 
\cdot (au_{\ast})^{2} \;, \qquad 
{\cal S}_{D} \left(= \dfrac{\pi a^{2}}{N^{2}L^{2}}\,S_{D} \right) 
\cong \dfrac{1}{2} 
\cdot (au_{\Lambda})^{2} \;.
\end{align}
In deriving this expression, we have utilized a formula 
\begin{align}
\label{212}
\int^{1}_{0}\dfrac{dt}{t^{\lambda}}\left(\dfrac{1}{\sqrt{1-t^{\kappa}}}-1\right)
- \dfrac{1}{\lambda-1}=\dfrac{\sqrt{\pi}}{\kappa}
\dfrac{\Gamma(\frac{1-\lambda}{\kappa})}
{\Gamma(\frac{\kappa-2\lambda+2}{2\kappa})} \;,
\end{align}
with $\lambda>1$. Notice that the finite part ${\cal S}_{U}$ is independent of the cutoff parameter $u_{\Lambda}$ and thus is universal quantity. Eliminating the parameter $u_{\ast}$ from (\ref{210}) and (\ref{211}), we find the relation between the universal part ${\cal S}_{U}$ and the ratio $l/a$ in the commutative regime: \footnotemark[2]\mbox{${}^{)}$}
\begin{align}
\label{213}
{\cal S}_{U} 
= -2\left\{\sqrt{\pi}\,\dfrac{\Gamma(\frac{2}{3})}{\Gamma(\frac{1}{6})}
\right\}^{3}\left(\dfrac{l}{a}\right)^{-2}\;. 
\end{align}
\footnotetext[2]{${}^{)}$ The dependence of (\ref{213}) on the noncommutativity parameter $a$ arises from the definition of the (dimensionless) entanglement entropy functional ${\cal S}_{U} \equiv (\pi a^{2}/N^{2}L^{2})\,S_{U}$.} 
Meanwhile, we find that the divergence part of the entanglement entropy ${\cal S}_{D}$ in the commutative regime is proportional to the area $L^{2}$:
\begin{align}
\label{214}
\dfrac{2\pi}{N^{2}}S_{D}
=L^{2} \cdot u_{\Lambda}^{2} \;.
\end{align}

In the deep ultraviolet bound $1/au_{\ast} \cong 0$, the ratio of the length $l$ to the noncommutativity parameter $a$ can be approximated as 
\begin{align}
\label{215}
\dfrac{l}{a} \cong \dfrac{\sqrt{\pi}}{3}\dfrac{\Gamma(\frac{1}{3})}{\Gamma(\frac{5}{6})} \cdot au_{\ast} \;.
\end{align}
Hereafter, we refer to the approximation in the deep ultraviolet bound as the {\it deep noncommutative regime} \cite{LBCF}. The entanglement entropy functional (\ref{208}) in the deep noncommutative regime can be divided into the universal part ${\cal S}_{U}$ and divergence part ${\cal S}_{D}$:
%
\begin{align}
\label{216}
{\cal S}_{U} 
= \dfrac{\sqrt{\pi}}{24}\dfrac{\Gamma(\frac{1}{3})}{\Gamma(\frac{5}{6})} \cdot (au_{\ast})^{4} \;, \qquad 
{\cal S}_{D} = \dfrac{1}{4} \cdot (au_{\Lambda})^{4} \;,
\end{align}
respectively. The relation between the finite part ${\cal S}_{U}$ and the ratio $l/a$ in the deep noncommutative regime is given by
\begin{align}
\label{217}
{\cal S}_{U}
=\dfrac{1}{8}\left\{\dfrac{3}{\sqrt{\pi}}\dfrac{\Gamma(\frac{5}{6})}{\Gamma(\frac{1}{3})}\right\}^{3}\left(\dfrac{l}{a}\right)^{4}\;.
\end{align}
We notice that the dependence of the finite part ${\cal S}_{U}$ on the ratio $l/a$ is quite different between the commutative regime and the deep noncommutative regime.  Eliminating the noncommutativity parameter $a$ from (\ref{215}) and  (\ref{216}), we find the relation between the divergence part $S_{D}$ and the length $l$ in the deep noncommutative regime:
\begin{align}
\label{218}
\dfrac{4\pi^{3/2}}{3N^{2}}\dfrac{\Gamma(\frac{1}{3})}{\Gamma(\frac{5}{6})} S_{D}
= L^{2}l \cdot \dfrac{u_{\Lambda}^{4}}{u_{\ast}} \;.
\end{align}
For $u_{\ast} \sim u_{\Lambda}$, this expression exhibits that the divergence part of the entanglement entropy $S_{D}$ in the deep noncommutative regime is proportional to the volume $L^{2}l$. The difference in the $l$-dependence between (\ref{214}) and (\ref{218}) would be understood as the {\it area}/{\it volume} law transition \cite{LBCF}.

%
%

\begin{figure}[H]
\centering
\hspace*{-15mm}
\begin{tabular}{cc}
\includegraphics[width=60mm]{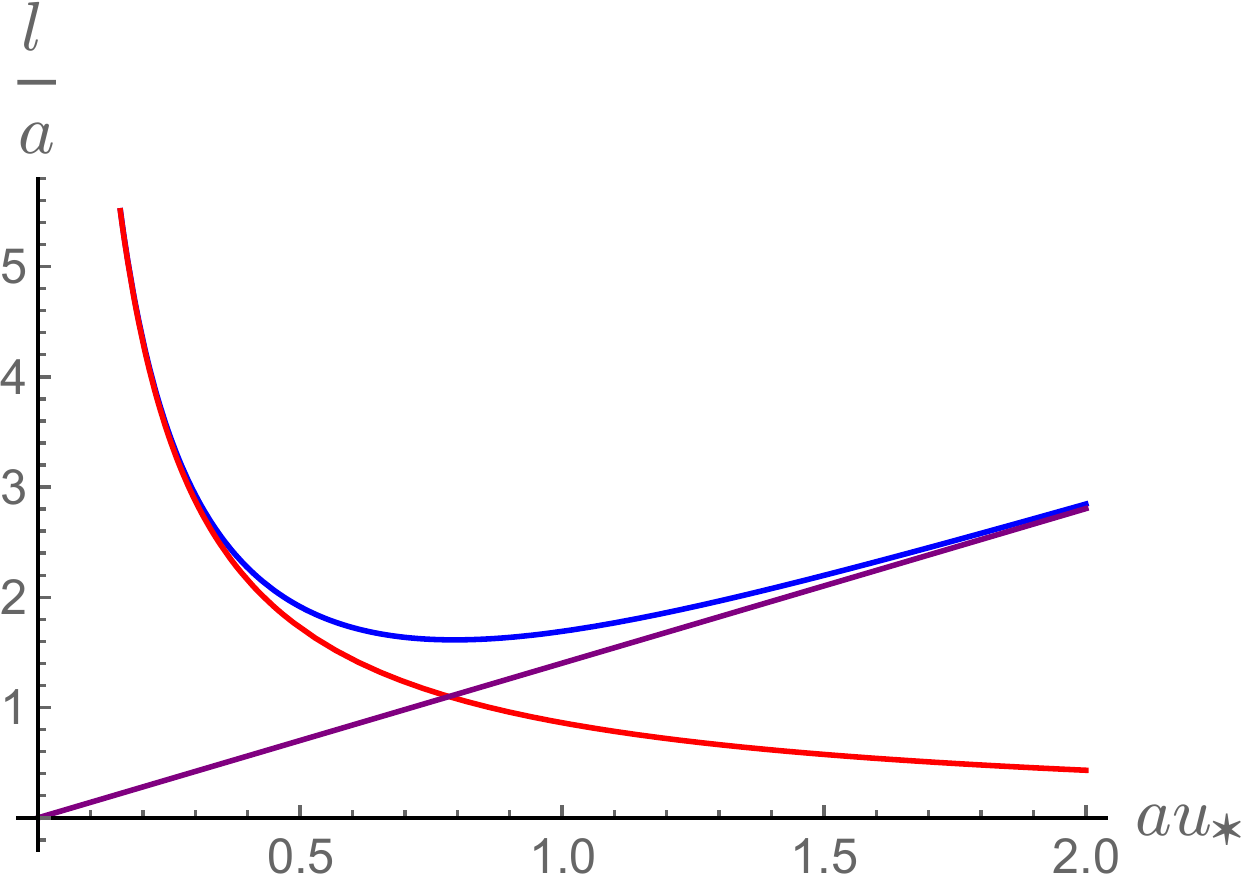} & 
\hspace*{5mm} \includegraphics[width=60mm]{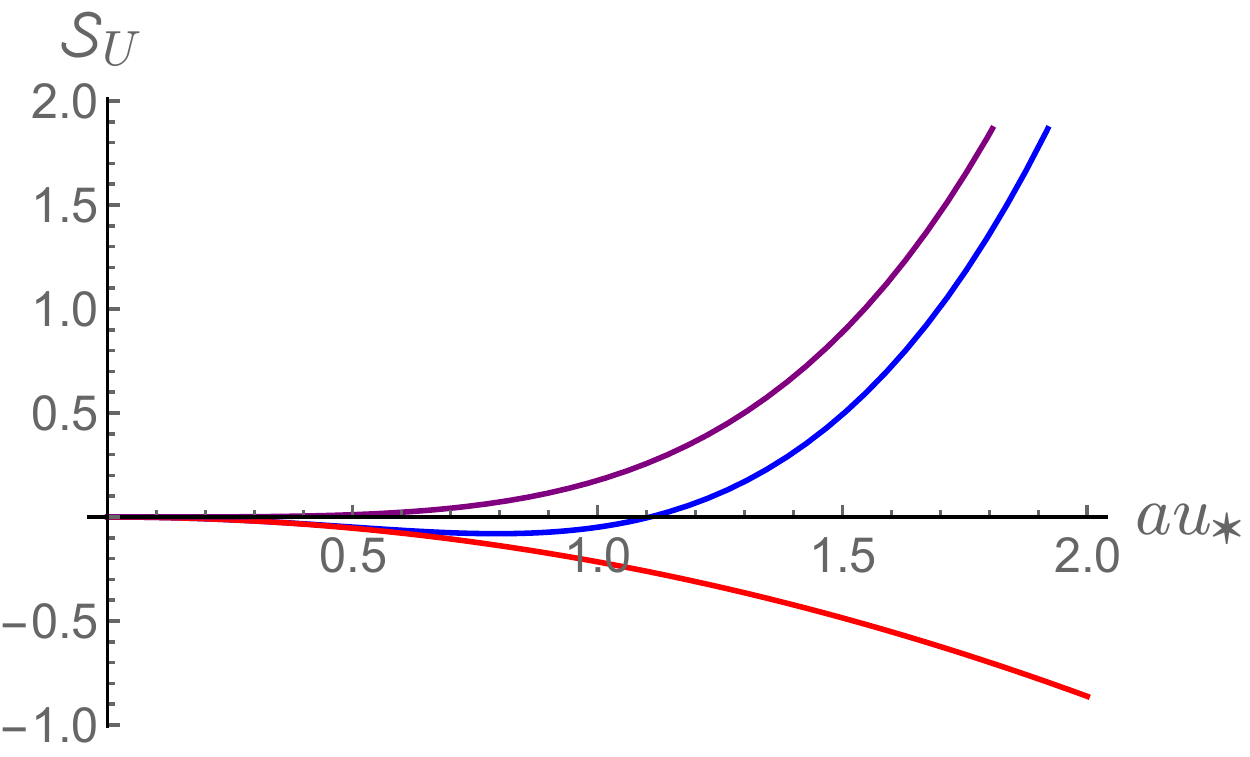} \\
(a) & (b) \\
\end{tabular}
\caption{(a)\,\,The variation with $au_{\ast}$ of the ratio $l/a$ . (b)\,\,The variation with $au_{\ast}$ of the universal part ${\cal S}_{U}\left(= \dfrac{\pi a^{2}}{N^{2}L^{2}}S_{U} \right)$. The blue curve line denotes the variation in the noncommutative theory. The red and purple curve lines denote the variation in the commutative regime and in the deep noncommutative regime, respectively.}
\label{F1}
\end{figure}

The variation with $au_{\ast}$ of $l/a$ is shown by Fig.\ref{F1}(a). The behavior of $l/a$ is quite different between the commutative regime and the deep noncommutative regime. The length $l$ has a minimum value $l_{\rm min} \sim 1.614\,a$ at $u_{\ast} \sim 0.7946/a$ in the noncommutative theory. The ratio $l/a$ is proportional to the inverse of $au_{\ast}$ for $u_{\ast} \ll 0.7946/a$ and is proportional to $au_{\ast}$ for $u_{\ast} \gg 0.7946/a$. The ratio $l/a$ increases in case of $au_{\ast} \to \infty$ as well as in case of $au_{\ast} \to 0$. This behavior is reminiscent of the UV/IR relation \cite{MRS}.

The dimensionless quantity (\ref{208}) can also be divided into the finite part ${\cal S}_{U}$ and divergence part ${\cal S}_{D}$:
\begin{subequations}
\begin{align}
\label{219a}
{\cal S}_{U} \left(=\dfrac{\pi a^{2}}{N^{2}L^{2}}S_{U} \right)
&= (au_{\ast})^{2}\int^{1}_{0} dt \, \dfrac{\sqrt{t^{4}+(au_{\ast})^{4}}}{t^{5}}
\left( \dfrac{1}{\sqrt{1-t^{6}}}-1 \right)  \nonumber \\
& -\dfrac{(au_{\ast})^{2}\sqrt{1+(au_{\ast})^{4}}}{4}
-\dfrac{1}{8} \ln
\left|\;\dfrac{1+\sqrt{1+\dfrac{1}{(au_{\ast})^{4}}}}
{1-\sqrt{1+\dfrac{1}{(au_{\ast})^{4}}}}\; \right| \;,  \\ 
\label{219b}
{\cal S}_{D} \left(=\dfrac{\pi a^{2}}{N^{2}L^{2}}S_{D} \right)
&= \dfrac{a^{2}u_{\Lambda}^{4}}{4u_{\ast}^{2}}
\sqrt{\left(\dfrac{u_{\ast}}{u_{\Lambda}}\right)^{4}+(au_{\ast})^{4}}
+ \dfrac{1}{8} \ln
\left|\;\dfrac{1+\sqrt{1+\dfrac{1}{(au_{\Lambda})^{4}}}}
{1-\sqrt{1+\dfrac{1}{(au_{\Lambda})^{4}}}}\; \right| \;.
\end{align}
\end{subequations}
In deriving the expressions above, we have utilized the formula
\begin{align}
\label{220}
\int dx \,\dfrac{\sqrt{x^{4}+k}}{x^{5}}
=-\dfrac{\sqrt{x^{4}+k}}{4x^{4}}-\dfrac{1}{8\sqrt{k}}
\ln\left|\;\dfrac{1+\sqrt{1+\dfrac{x^{4}}{k}}}{1-\sqrt{1+\dfrac{x^{4}}{k}}}\; 
\right| \;, 
\end{align}
with a constant $k$. The divergence part ${\cal S}_{D}$ becomes $0$ when the parameter $u_{\Lambda}$ is taken to be $0$. The universal part ${\cal S}_{U}$ given by (\ref{219a}) is also a function of the dimensionless quantity $au_{\ast}$. The variation with $au_{\ast}$ of ${\cal S}_{U}$ is shown by Fig.\ref{F1}(b). The behavior of ${\cal S}_{U}$ is also different between the commutative regime and the deep noncommutative regime. The universal part ${\cal S}_{U}$ given by (\ref{219a}) takes a minimum value at $u_{\ast} \sim 0.7946/a$. This is the same value at which the length $l$ takes the minimum value $l_{\rm min}$ in the noncommutative theory.

\begin{figure}[H]
\centering
\hspace*{0mm}
\includegraphics[width=80mm]{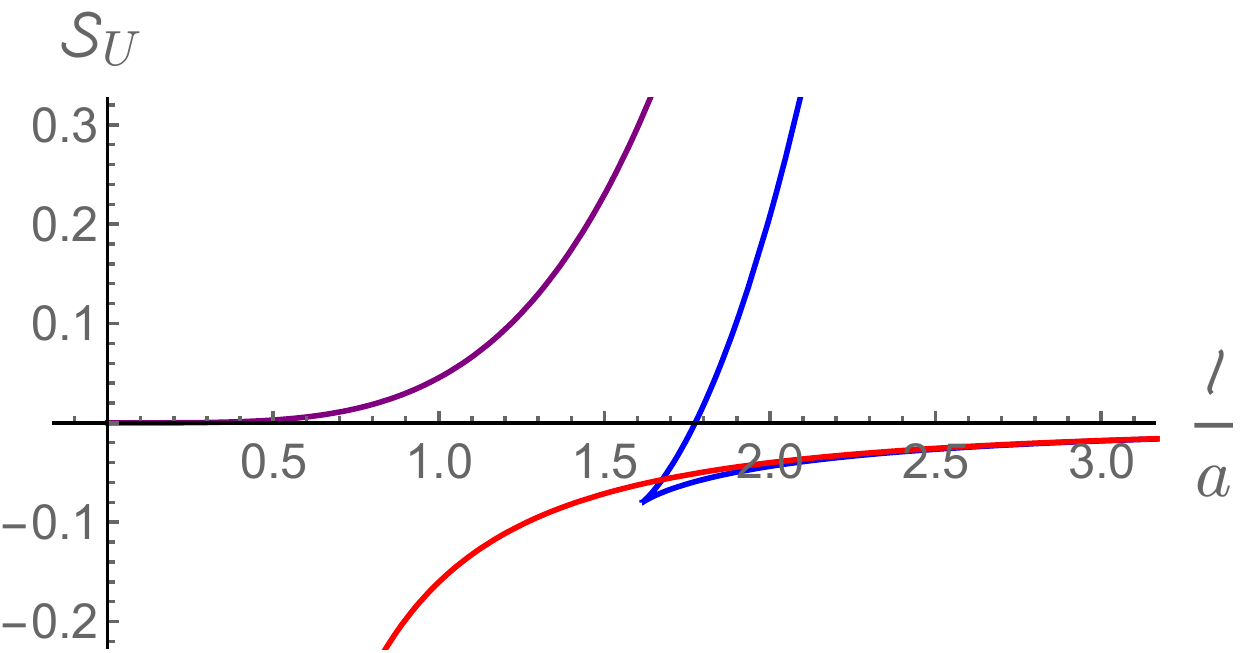} 
\caption{The variation with $l/a$ of ${\cal S}_{U}$ 
. The blue curve line denotes the variation in the noncommutative theory. The red and purple curve lines denote the variation in the commutative regime and in the deep 
noncommutative regime, respectively.}
\label{F2}
\end{figure}

We can evaluate the dependence of the universal part ${\cal S}_{U}$ 
on the ratio $l/a$ 
numerically. The variation with $l/a$ of the universal part ${\cal S}_{U}$ is shown by Fig.\ref{F2}. 

Although the sign of the universal term ${\cal S}_{U}$ can take negative, the sign of the entanglement entropy is positive. As we will see later, the sign of the derivative of the universal term ${\cal S}_{U}$ is more important.

There is major difference in the dependence of the universal part ${\cal S}_{U}$ on the ratio $l/a$ between in the commutative regime and in the deep noncommutative regime. The curve of ${\cal S}_{U}$ is concave downward in the commutative regime (shown as a red curve line in Fig.\ref{F2}), while that is concave upward in the deep noncommutative regime (shown as a purple curve line in Fig.\ref{F2}). This concave upward curve suggests that the behavior of ${\cal S}_{U}$ in the deep noncommutative regime becomes unphysical. In the noncommutative theory, the curve of ${\cal S}_{U}$ branches into a concave downward curve and a concave upward curve at $l/a=l_{min}/a$ (shown as blue curve lines in Fig.\ref{F2}). Fig.\ref{F2} shows that the blue concave downward curve (lower branch) and the blue concave upward curve (upper branch) asymptotically approach the red concave downward curve and the purple concave upward curve in the limit $l/a \to \infty$, respectively.

We see that the universal part ${\cal S}_{U}$ with identical the ratio $l/a (> l_{min}/a)$ actually have different value in the noncommutative theory. Therefore, the concave downward curve becomes presumably dominated for the curve of ${\cal S}_{U}$ in the noncommutative theory. When the concave upward curve of ${\cal S}_{U}$ is realized in the noncommutative theory, the derivative of the universal part ${\cal S}_{U}$ with respect to the ratio $l/a$ seem to be discontinuous at the point $l/a = l_{min}/a$. This behavior could be interpreted as the {\it area}/{\it volume} law transition \cite{LBCF} from the viewpoint of the universal part ${\cal S}_{U}$.

%
%
\section{Strong subadditivity and Mutual information} 

\setcounter{equation}{0}
\addtocounter{enumi}{1}

\begin{figure}[H]
\centering
\hspace*{0mm}
\includegraphics[width=100mm]{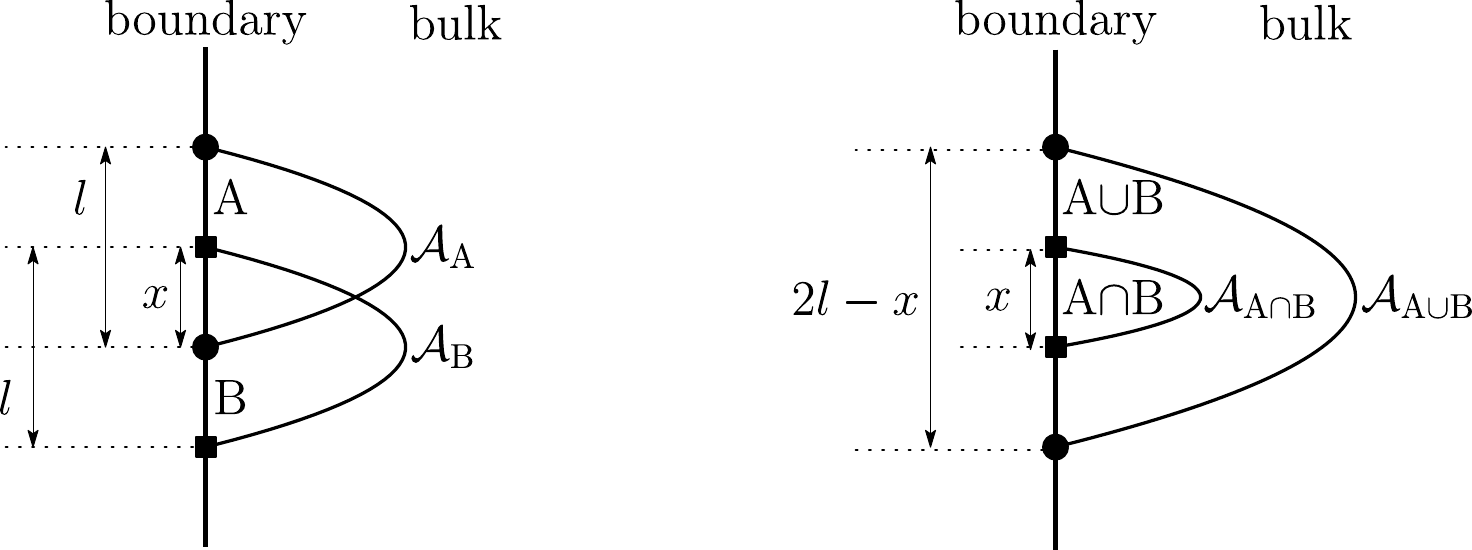} 
\caption{(Left) Two overlapping regions A and B of the boundary, with their respective minimal bulk hypersurfaces ${\cal A}_{\rm A}$ and ${\cal A}_{\rm B}$. (Right) Two regions A$\cup$B and A$\cap$B of the boundary, with their respective minimal hypersurfaces ${\cal A}_{{\rm A}\cup{\rm B}}$ and ${\cal A}_{{\rm A}\cap{\rm B}}$.}
\label{F3}
\end{figure}

Entanglement entropies are subject to an inequality known as strong subadditivity \cite{EHLMBR1, EHLMBR2}. This inequality can be stated as $S(A) + S(B) \geq S(A \cup B) + S(A \cap B)$ for any two regions of space $A$ and $B$. We would like to examine if the holographic entanglement entropies in the noncommutative Yang-Mills theory are subject to the strong subadditivity. We consider two infinite strip each of length $l$ overlapped with a width $x$ along the $x^{2}$-direction. A conceptual figure is given in Fig.\ref{F3}.

\begin{figure}[H]
\centering
\hspace*{-15mm}
\begin{tabular}{cc}
\includegraphics[width=60mm]{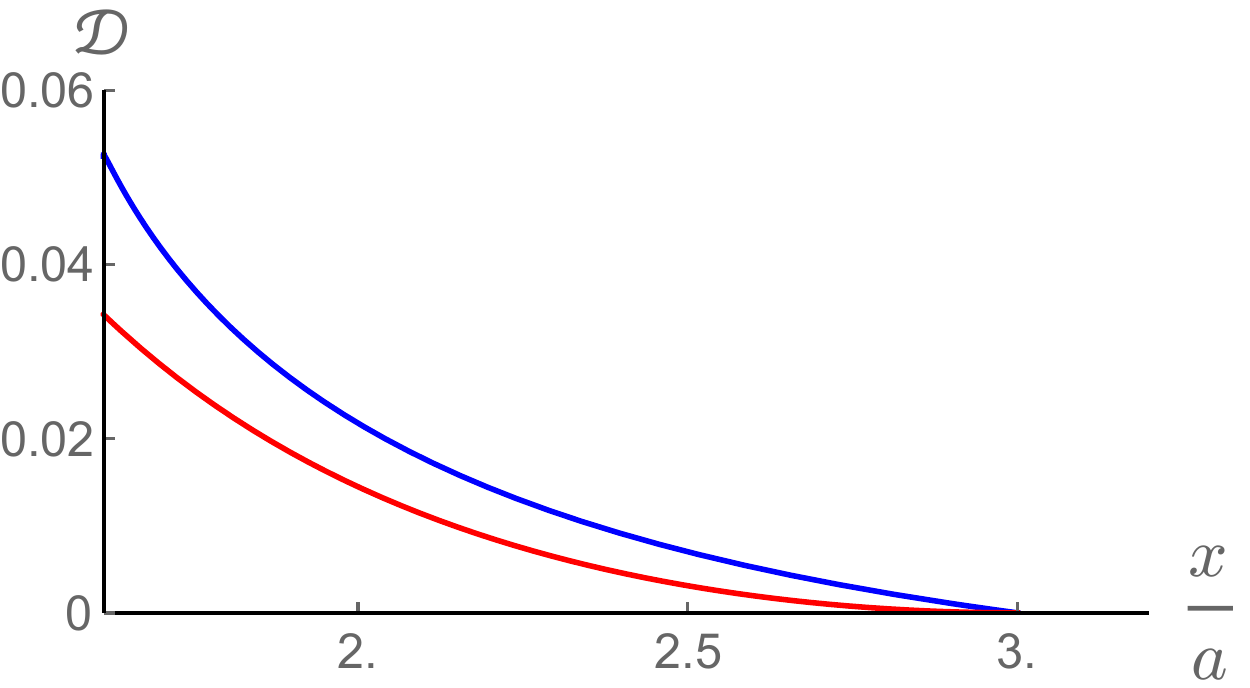} & 
\hspace*{5mm} \includegraphics[width=60mm]{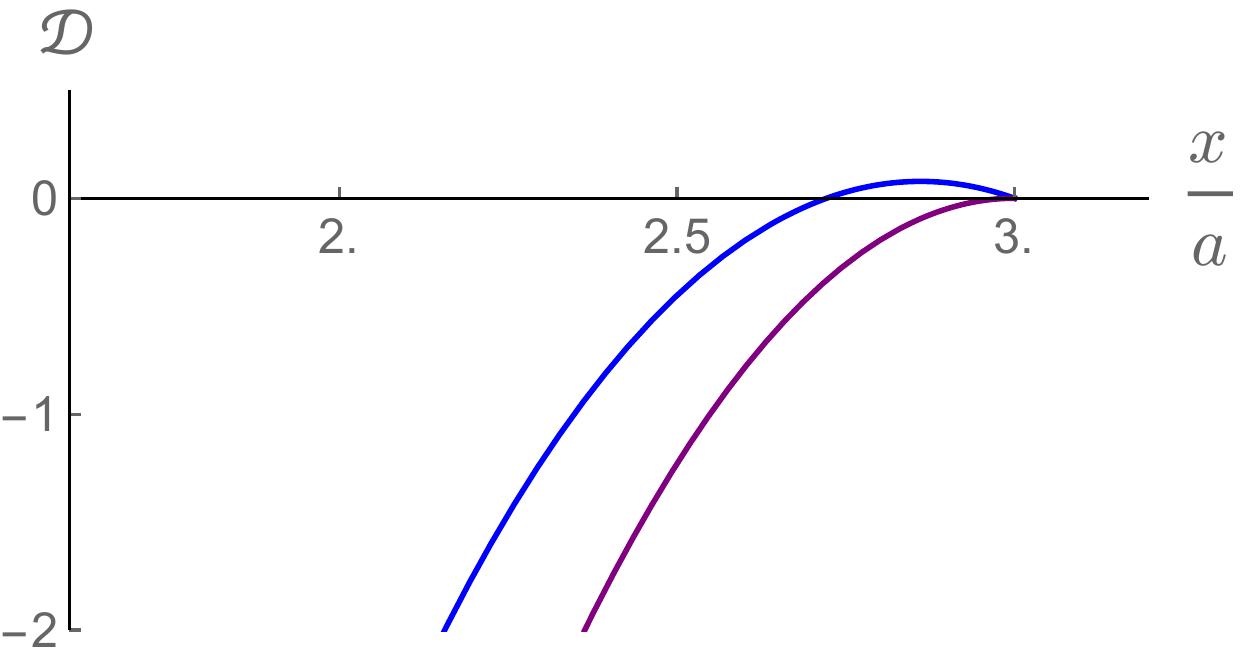} \\
(a) & (b) \\
\end{tabular}
\caption{(a) The variation with $x/a$ of the quantity ${\cal D}$ for $l/a=3$ and $u_{\ast} < 0.7946/a$. (b) The variation of the quantity ${\cal D}$ with $x/a$ for $l/a=3$ and $u_{\ast} > 0.7946/a$. The blue curve line denotes the variation in the noncommutative theory. The red and purple curve lines denote the variation in the commutative regime and in the deep noncommutative regime, respectively.}
\label{F4}
\end{figure}

Let us define the following quantity
\begin{align}
\label{301}
D(x)= 2S_{U} \Bigl( \dfrac{l}{a} \Bigr) 
- S_{U}\Bigl( \dfrac{x}{a} \Bigr) - S_{U}\Bigl( \dfrac{2l-x}{a} \Bigr) \;.
\end{align}
The quantity ${\cal D}=\dfrac{\pi a^{2}}{N^{2}L^{2}}D$ becomes positive when the strong subadditivity inequality is satisfied. The variation of $d$ with $x/a$ is shown by Fig.\ref{F4}. 

As we expected, the strong subadditivity inequality for the entanglement entropies in the commutative regime is satisfied. On the other hand, the strong subadditivity inequality for the entanglement entropies in the deep noncommutative regime is not satisfied. The strong subadditivity inequality for the entanglement entropies in the noncommutative theory is also satisfied under the condition of $u_{\ast} < 0.7946/a$. It is interesting to note that the strong subadditivity inequality for the entanglement entropies in the noncommutative theory is partially satisfied under the condition of $u_{\ast} > 0.7946/a$, in spite of their unphysical behavior. We find that the inequality in the noncommutative theory is completely satisfied when the two subsystems almost overlap. \\

\begin{figure}[H]
\centering
\hspace*{0mm}
\includegraphics[width=100mm]{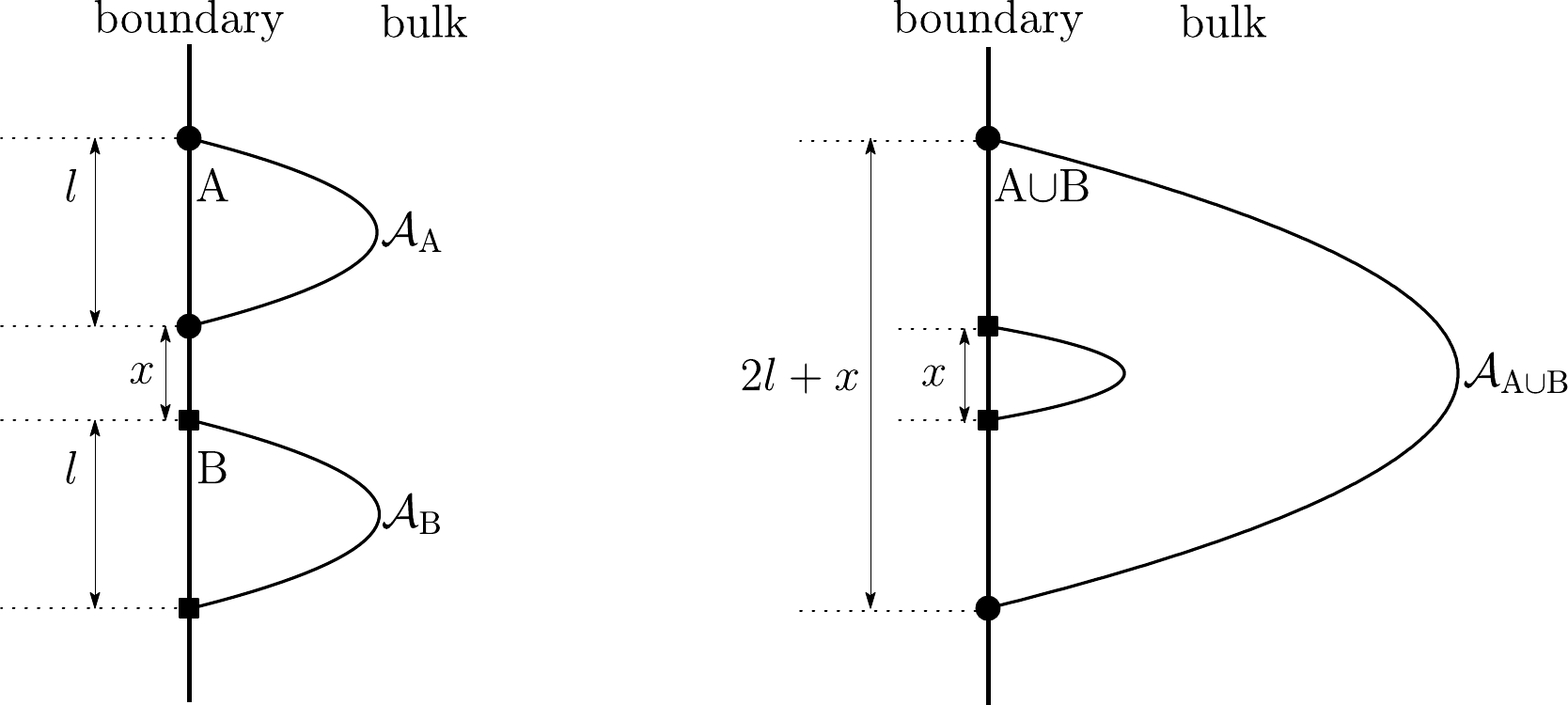} 
\caption{(Left) Two separated regions A and B of the boundary, with their respective minimal bulk hypersurfaces ${\cal A}_{\rm A}$ and ${\cal A}_{\rm B}$. (Right) The region A$\cup$B of the boundary, with their respective minimal hypersurface ${\cal A}_{{\rm A}\cup{\rm B}}$.}
\label{F5}
\end{figure}

The mutual information $I(A,\;B)$ of any two regions of space $A$ and $B$ can be written by the entanglement entropies as $I(A,\;B)=S(A)+S(B)-S(A \cup B)$. It measures how much we learn about one region of space $A$ by observing the other region of space $B$. This interpretation convinces us that the mutual information $I(A,\;B)$ have to be nonnegative. The nonnegativity property of the mutual information is guaranteed by the subadditivity of the entanglement entropy. We would like to examine if the mutual information in the noncommutative Yang-Mills theory are subject to the nonnegativity property. We consider two infinite strip each of length $l$ separated by a distance $x$ along the $x^{2}$-direction. A conceptual figure is given in Fig.\ref{F5}.

\begin{figure}[H]
\centering
\hspace*{0mm}
\includegraphics[width=80mm]{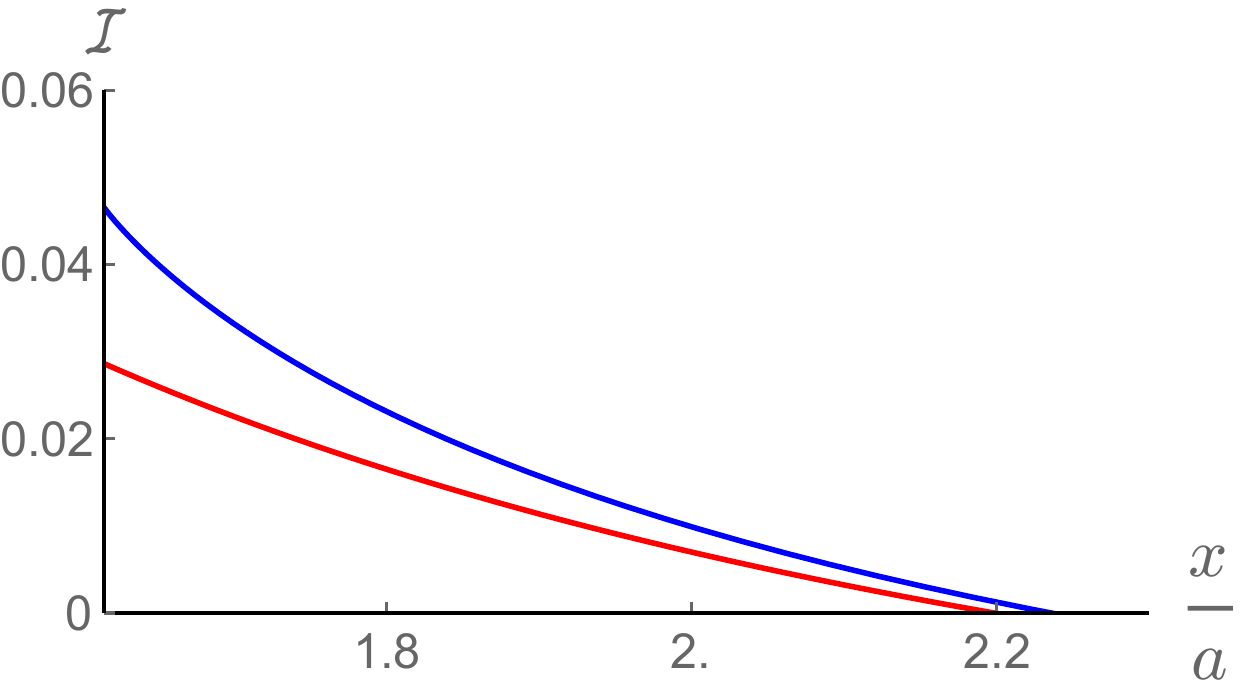} 
\caption{The variation with $x/a$ of the quantity ${\cal I}$ for $l/a=3$ and $u_{\ast} < 0.7946/a$. The blue curve line denotes the variation in the noncommutative theory. The red curve line denotes the variation in the commutative regime.}
\label{F6}
\end{figure}

Let us define the following quantity
\begin{align}
\label{302}
I(x)= 2S_{U} \Bigl( \dfrac{l}{a} \Bigr) 
- S_{U}\Bigl( \dfrac{x}{a} \Bigr) - S_{U}\Bigl( \dfrac{2l+x}{a} \Bigr) \;.
\end{align}
The variation of the quantity ${\cal I}=\dfrac{\pi a^{2}}{N^{2}L^{2}}I$ with $x/a$ is shown by Fig.\ref{F6}. 

We find that the value of the mutual information in the noncommutative theory under the condition of $u_{\ast} < 0.7946/a$ is slightly larger than that in the commutative regime. The mutual information in the noncommutative theory under the condition of $u_{\ast} > 0.7946/a$ (and also in the deep noncommutative regime) takes the negative value for all values of the distance $x$. 

%
%
\section{Entropic c-function} 

\setcounter{equation}{0}
\addtocounter{enumi}{1}

\begin{figure}[H]
\centering
\hspace*{0mm}
\includegraphics[width=80mm]{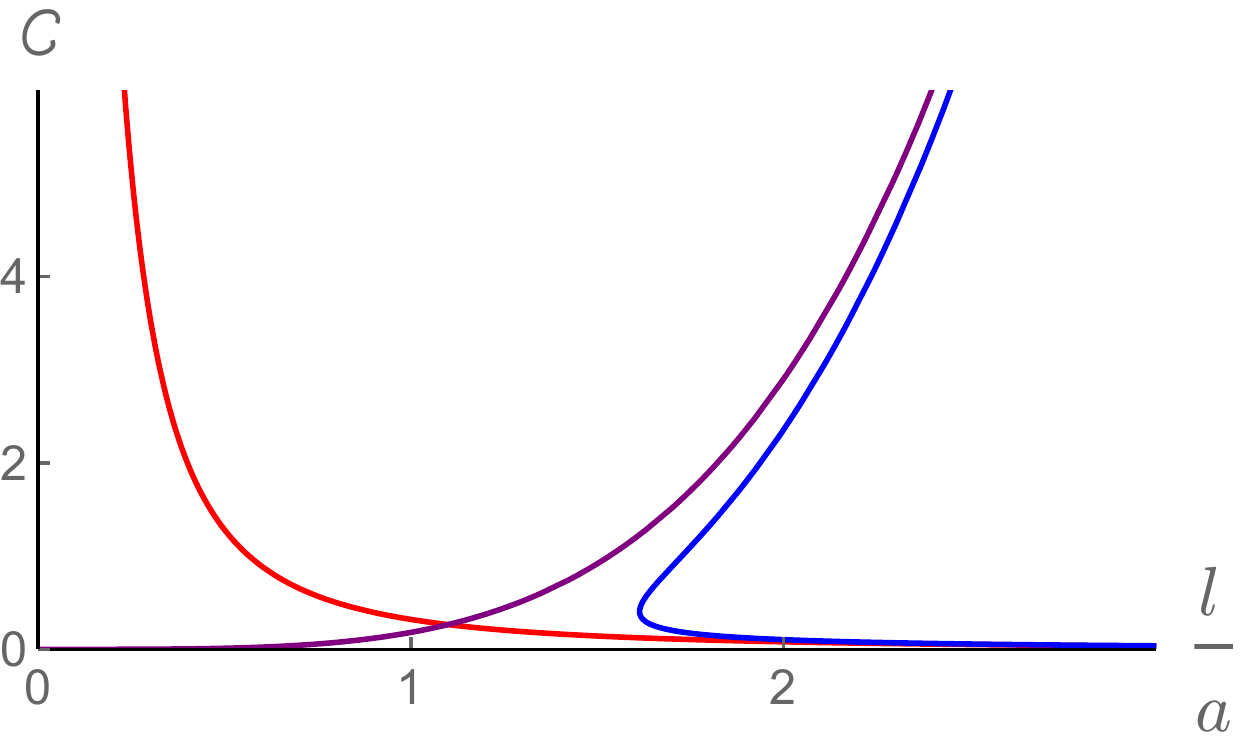} 
\caption{The variation with $l/a$ of ${\cal C}$. The blue curve line denotes the variation in the noncommutative theory. The red and purple curve lines denote the variation in the commutative regime and in the deep noncommutative regime, respectively.}
\label{F7}
\end{figure}

It is well known that there exists a so called c-function that is a positive real function and is monotonically decreasing under the renormalization group (RG) flow. The $c$-function can be defined by means of the entanglement entropy, and that is called entropic $c$-function \cite{HCMH, HCCDFMH, HCMH2}. For the infinite strip subsystem with the length $l$, the entropic $c$-function denoted by $C$ can be rewritten as \cite{SRTT, IRKDKAM, LBCF} 
\begin{align}
\label{401}
C(l)=\dfrac{dS_{A}}{d\,\ln l}=l\dfrac{dS_{A}}{dl}\;. 
\end{align}
Note that this quantity does not depend on the cutoff parameter $u_{\Lambda}$. The variation of ${\cal C}(l)=\dfrac{\pi a^{2}}{N^{2}L^{2}}C(l)$ with $l/a$ is shown by Fig.\ref{F7}. 

Since the $c$-function denoted by ${\cal C}(l)$ measures the number of degrees of freedom, it is expected to satisfy the inequality ${\cal C}(l_{\rm UV}) \geq {\cal C}(l_{\rm IR})\;(\text{for}\;l_{\rm UV} \leq l_{\rm IR})$ and the derivative of ${\cal C}$ with respect to $l$ is expected to be negative. As we expected, the entropic $c$-function in the commutative regime (shown as a red curve line in Fig.\ref{F7}) satisfies such properties. The entropic $c$-function in the deep noncommutative regime (shown as a purple curve line in Fig.\ref{F7}), however, does not follow the expected behavior. It diverges as $l$ approaches to $\infty$. The entropic $c$-function in the noncommutative theory (shown as a blue curve line in Fig.\ref{F7}) under the condition of $u_{\ast} < 0.7946/a$ also satisfies the inequality ${\cal C}(l_{\rm UV}) \geq {\cal C}(l_{\rm IR})\;(\text{for}\;l_{\rm UV} \leq l_{\rm IR})$ and the derivative of ${\cal C}$ with respect to $l$ is negative. The turning point on the blue curves in Fig.\ref{F2} and Fig.\ref{F7} can be observed at the same value for $l/a\,(=l_{min}/a)$. 

The behavior of the entropic $c$-function in the commutative regime under the condition of $l \to 0$ is similar to that in the deep noncommutative regime under the condition of $l \to \infty$. This phenomenon also seems to be relevance to the UV/IR mixing phenomenon \cite{MRS}, in the sense that UV divergence of the commutative regime appears to be replaced by IR singularity of the deep noncommutative regime. 

%
%
\section{Finite temperature} 

\setcounter{equation}{0}
\addtocounter{enumi}{1}

In this section, we consider the holographic entanglement entropy in the noncommutative Yang-Mills theory at finite temperature.  A holographic description of the noncommutative Yang-Mills theory at finite temperature is given by
\begin{align}
\label{501}
ds^{2} &=R^{2}
\Bigl[u^{2} \bigl\{f(u)dx_{0}{}^{2}+dx_{1}{}^{2} + h(u)(dx_{2}{}^{2}
+dx_{3}{}^{2}) \bigr\} 
+ \left(\dfrac{du^{2}}{u^{2}f(u)}+d\Omega_{5}{}^{2} \right) \Bigr] \,, \nonumber \\[-3mm]
\quad \\[-3mm]
f& =1-\left(\dfrac{u_{T}}{u} \right)^{4} \;,
\nonumber 
\end{align}
where $u_{T}$ is a parameter with dimension of mass, and is the lower bound of $u$. The corresponding temperature $T$ of the background can be obtained to be $T=u_{T}/\pi$. 

We compute the entanglement entropy for an infinite strip specified by (\ref{203}) with $L \to \infty$. Under this configuration, the entanglement entropy denoted by $S_{AT}$ at the stable solution is modified to include the parameter $u_{T}$:
\begin{align}
\label{502}
S_{AT} = \dfrac{{\cal A}}{4G_{N}^{(10)}}
= \dfrac{N^{2}L^{2}}{\pi}\int^{u_{\Lambda}}_{u_{\ast}} 
du\;u^{6} \sqrt{\dfrac{1+(au)^{4}}{(u^{6}-u_{\ast}^{6})(u^{4}-u_{T}^{4})}} \;.
\end{align}
The length $l$ is also modified to include the parameter $u_{T}$:
\begin{align}
\label{503}
\dfrac{l}{2} = u_{\ast}^{2} \int^{\infty}_{u_{\ast}} du\;
\sqrt{\dfrac{1+(au)^{4}}{(u^{6}-u_{\ast}^{6})(u^{4}-u_{T}^{4})}} \;.
\end{align}
We can find that in the large $l$ limit, the main contribution of (the finite part of) the integrals (\ref{502}) and (\ref{503}) coming from the region near $u \sim u_{\ast} \sim u_{T}$ leads to the relation, 
\begin{align}
\label{504}
S_{AT}^{finite} = \dfrac{\pi^{2}}{2} N^{2}T^{3} \cdot L^{2}l \;.
\end{align}

The entanglement entropy given by (\ref{504}) is proportional to the volume $L^{2}l$. Thus it is extensive as in the thermal entropy. 
The $l$-dependence of the entanglement entropy shown by the expression (\ref{504}) is at first glance similar to that shown by the expression (\ref{218}). It should be noted, however, that the expression (\ref{218}) represents the property of the divergent part of the entanglement entropy, while equation  (\ref{504}) represents the property of the finite part of the entanglement entropy.

The (dimensionless) entanglement entropy functional ${\cal S}_{AT} = (\pi a^{2}/N^{2}L^{2})\,S_{AT}$ can be rewritten as 
\begin{align}
\label{505}
{\cal S}_{AT} = (au_{\ast})^{2}\int^{1}_{u_{\ast}/u_{\Lambda}} \dfrac{dt}{t^{5}}
\sqrt{\dfrac{t^{4}+(au_{\ast})^{4}}{(1-t^{6})
\left( 1- \left(\dfrac{\tau t}{au_{\ast}} \right)^{4} \right)}} \,
\end{align}
where $t \equiv u_{\ast}/u$ and $\tau$ is a dimensionless parameter defined
by $\tau \equiv au_{T}$. The ratio of the length $l$ to the noncommutativity parameter $a$ is also modified to include the parameter $\tau$:
\begin{align}
\label{506}
\dfrac{l}{a} 
= \dfrac{2}{au_{\ast}}
\int^{1}_{0}dt\; t\,
\sqrt{\dfrac{t^{4}+(au_{\ast})^{4}}{(1-t^{6})
\left( 1- \left(\dfrac{\tau t}{au_{\ast}} \right)^{4} \right)}}\;.
\end{align}

The relation (\ref{504}) can also be rewritten in terms of ${\cal S}_{AT}^{finite}=(\pi a^{2}/N^{2}L^{2})\,S_{AT}^{finite}, \;l/a$ and $\tau$:
\begin{align}
\label{507}
{\cal S}_{AT}^{finite} = \tau^{3} \cdot \dfrac{l}{a}\;.
\end{align}

The entanglement entropy functional given by (\ref{505}) can be divided into the universal part ${\cal S}_{UT}$ and divergence part ${\cal S}_{DT}$:
\begin{subequations}
\begin{align}
\label{508a}
{\cal S}_{UT} 
&= (au_{\ast})^{2}\int^{1}_{0} dt \, \dfrac{\sqrt{t^{4}+(au_{\ast})^{4}}}{t^{5}}
\Biggl\{ \dfrac{1}{\sqrt{(1-t^{6})
\left( 1- \left(\dfrac{\tau t}{au_{\ast}} \right)^{4} \right)}}-1 \Biggr\}  
\nonumber \\
& -\dfrac{(au_{\ast})^{2}\sqrt{1+(au_{\ast})^{4}}}{4}
- \dfrac{1}{8} \ln
\left|\;\dfrac{1+\sqrt{1+\dfrac{1}{(au_{\ast})^{4}}}}
{1-\sqrt{1+\dfrac{1}{(au_{\ast})^{4}}}}\; \right| \;, \\ 
\label{508b}
{\cal S}_{DT} 
&= \dfrac{a^{2}u_{\Lambda}^{4}}{4u_{\ast}^{2}}
\sqrt{\left(\dfrac{u_{\ast}}{u_{\Lambda}}\right)^{4}+(au_{\ast})^{4}}
+ \dfrac{1}{8}\ln
\left|\;\dfrac{1+\sqrt{1+\dfrac{1}{(au_{\Lambda})^{4}}}}
{1-\sqrt{1+\dfrac{1}{(au_{\Lambda})^{4}}}}\; \right| \;.
\end{align}
\end{subequations}

Although the universal part ${\cal S}_{UT}$ depends on the parameter $\tau$, related to the temperature $T$, the divergence part ${\cal S}_{DT}$ does not depend on the parameter $\tau$. This fact indicates that the property of the universal part of the entanglement entropy is modified at finite temperature, while the property of the divergent part of the entanglement entropy is not modified at all.

\begin{figure}[H]
\centering
\hspace*{-15mm}
\begin{tabular}{cc}
\includegraphics[width=60mm]{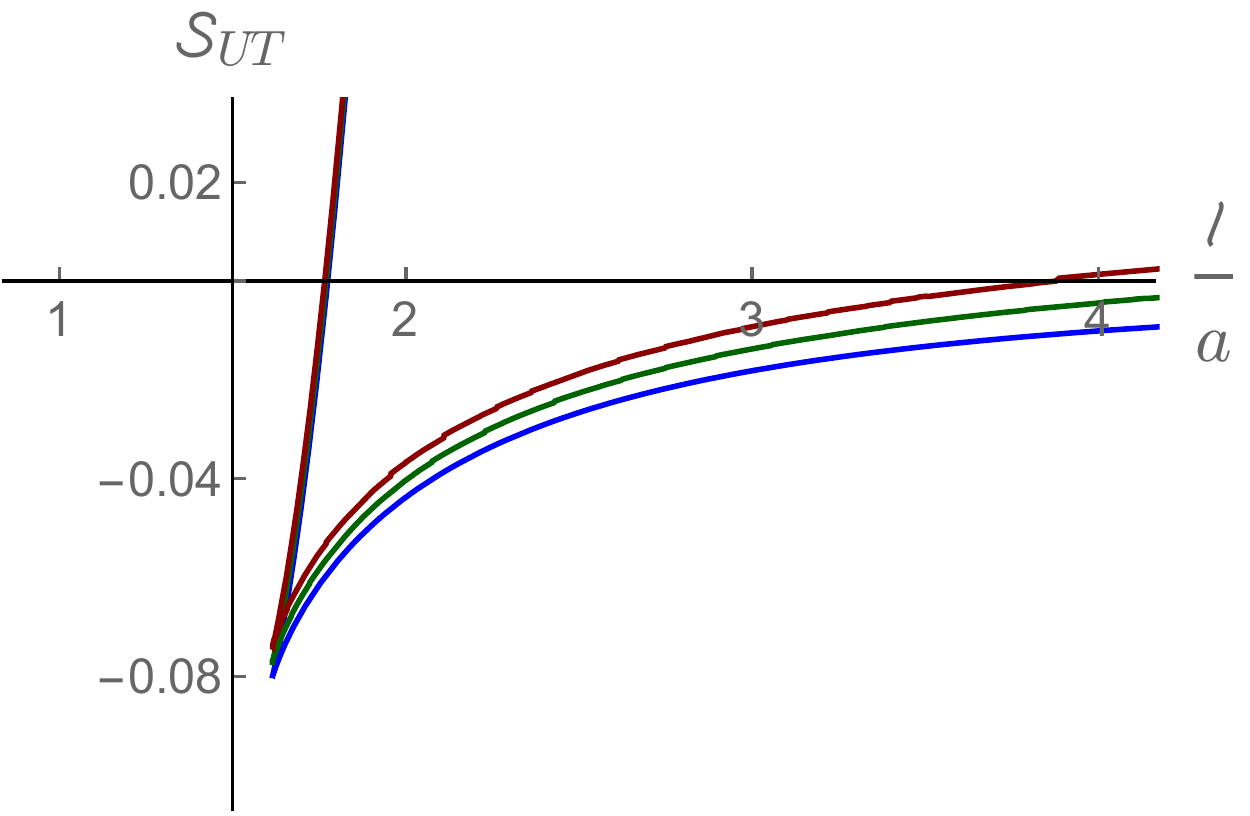} & 
\hspace*{5mm} \includegraphics[width=60mm]{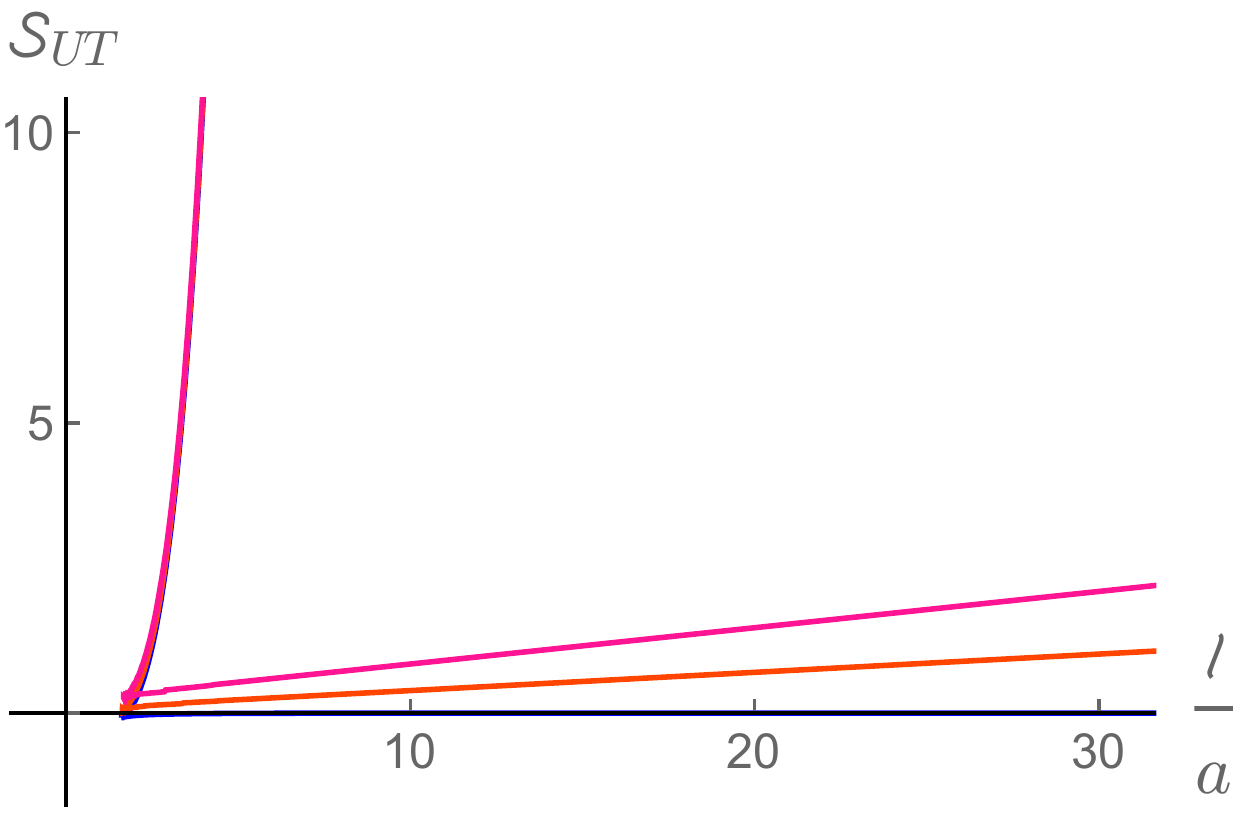} \\
(a) & (b) \\
\end{tabular}
\caption{The variation with $l/a$ of ${\cal S}_{UT}$ 
in the noncommutative theory. (a) The blue, green and brown curve lines correspond respectively to $\tau=0.00,\;0.16$ and $0.19$. (b) The blue, orange and magenta curve lines correspond respectively to $\tau=0.00,\;0.40$ and $0.50$.}

\label{F8}
\end{figure}

We can also evaluate the dependence of the universal part of the entanglement entropy functional $S_{UT}$ given by (\ref{508a}) on the length $l$ given by (\ref{506}) numerically. The variation with $l/a$ of the universal part ${\cal S}_{UT}$ is shown by Fig.\ref{F8}. 

Here, $\tau=0$ corresponds to the zero temperature case discussed in the previous sections. (It should be noted that the domain and range of the graph in Fig. \ref{F8} are different from that in Fig. \ref{F2}.)

Notice that the minimum length $l_{min}$ exists even at finite temperature. It can be found that the value of $l_{min}$ increases with increasing temperature $\tau$. The change in the minimum length $l_{min}$ with temperature $\tau$ is however slight. There are no significant changes in the $l$-dependence of the entanglement entropy in the region of $au_{\ast} > \tau$ ($u_{\ast} > u_{T}$). In contrast, Fig.\ref{F8}(b) shows that the relationship expressed in (\ref{507}) is approximately satisfied between ${\cal S}_{UT}$ and $l/a$ in the large $l$ limit (in the region of $au_{\ast} \sim \tau$ or $u_{\ast} \sim u_{T}$).

%
%

Generally, there is another surface that satisfy the boundary condition (\ref{206}) because of the holographic dual of the field theory at finite temperature involves black hole horizons \cite{IBAFLAPZCATE}. 
The surface is parametrized as 
\begin{align}
\label{509}
y=x_{2}=\pm \dfrac{l}{2}, \qquad u=u_{T}\;.
\end{align}

\begin{figure}[H]
\centering
\hspace*{0mm}
\includegraphics[width=80mm]{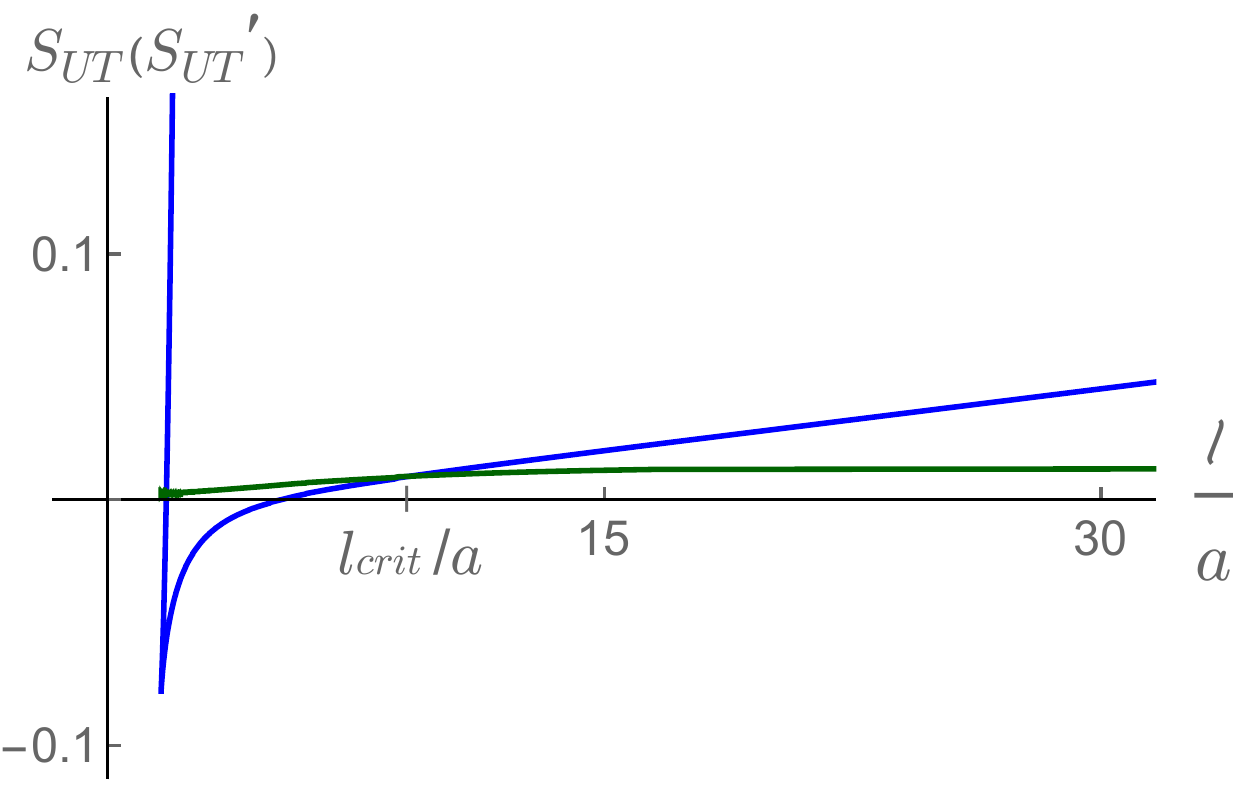} 
\caption{The variation with $l/a$ of ${\cal S}_{UT}$(blue curve line) and ${\cal S}_{UT}'$(green curve line) in the noncommutative theory. Both blue and green curve lines correspond to $\tau=0.15$.}
\label{F9}
\end{figure}

We call the surface that is parameterized as (\ref{509}) {\it piecewise} smooth surface, to distinguish it from the smooth minimal surface. There are two candidates of the surfaces to which Ryu-Takayanagi prescription can be applied. Let us compute the area of the piecewise smooth surface denoted by ${\cal A}'$ and examine the behavior of two areas ${\cal A}$ and ${\cal A}'$ as a function of the ratio $l/a$. The induced line elements for different segments are
\begin{subequations}
\begin{align}
\label{510a}
\dfrac{ds^{2}}{R^{2}} &=
u^{2} \bigl\{f(u)dx_{0}{}^{2}+dx_{1}{}^{2} + h(u)dx_{3}{}^{2} \bigr\} 
+ \dfrac{du^{2}}{u^{2}f(u)} + d\Omega_{5}{}^{2} 
\quad \text{for} \quad  y=\pm \dfrac{l}{2} \,, \quad \\[2mm]
\label{510b}
\dfrac{ds^{2}}{R^{2}} &= 
u^{2} \bigl\{f(u)dx_{0}{}^{2}+dx_{1}{}^{2} 
+ h(u_{T})(dx_{2}{}^{2} + dx_{3}{}^{2} \bigr\} 
+ d\Omega_{5}{}^{2} 
\quad \text{for} \quad \text{for} \;\;u=u_{T} \,.
\end{align}
\end{subequations}
The entanglement entropy denoted by $S_{AT}'$ is then given by:
\begin{align}
\label{511}
S_{AT}' = \dfrac{{\cal A}'}{4G_{N}^{(10)}}
= \dfrac{N^{2}L^{2}}{\pi}
\left\{ \int^{u_{\Lambda}}_{u_{\ast}} 
du\;u^{3} \sqrt{\dfrac{1+(au)^{4}}{u^{4}-u_{T}^{4}}} 
+ \dfrac{l}{2}u_{T}^{3} \right\} \;.
\end{align}
The (dimensionless) entanglement entropy functional ${\cal S}_{AT}' = (\pi a^{2}/N^{2}L^{2})\,S_{AT}'$ can also be divided into the universal part ${\cal S}_{UT}'$ and divergence part ${\cal S}_{DT}'$:
\begin{subequations}
\begin{align}
\label{512a}
{\cal S}_{UT}' 
&= (au_{\ast})^{2}\int^{1}_{0} dt \, \dfrac{\sqrt{t^{4}+(au_{\ast})^{4}}}{t^{5}}
\Biggl\{ \dfrac{1}{\sqrt{1- \left(\dfrac{\tau t}{au_{\ast}} \right)^{4}}}-1 
\Biggr\} \nonumber \\
& + \dfrac{\tau^{3}}{au_{\ast}}
\int^{1}_{0}dt\; t\,
\sqrt{\dfrac{t^{4}+(au_{\ast})^{4}}{(1-t^{6})
\left( 1- \left(\dfrac{\tau t}{au_{\ast}} \right)^{4} \right)}} \nonumber \\
& -\dfrac{(au_{\ast})^{2}\sqrt{1+(au_{\ast})^{4}}}{4}
- \dfrac{1}{8} \ln
\left|\;\dfrac{1+\sqrt{1+\dfrac{1}{(au_{\ast})^{4}}}}
{1-\sqrt{1+\dfrac{1}{(au_{\ast})^{4}}}}\; \right| \;,  \\ 
\label{512b}
{\cal S}_{DT}'
&= \dfrac{a^{2}u_{\Lambda}^{4}}{4u_{\ast}^{2}}
\sqrt{\left(\dfrac{u_{\ast}}{u_{\Lambda}}\right)^{4}+(au_{\ast})^{4}}
+ \dfrac{1}{8}\ln
\left|\;\dfrac{1+\sqrt{1+\dfrac{1}{(au_{\Lambda})^{4}}}}
{1-\sqrt{1+\dfrac{1}{(au_{\Lambda})^{4}}}}\; \right| \;.
\end{align}
\end{subequations}
The divergence part ${\cal S}_{DT}'$ is the same as ${\cal S}_{DT}$, and thus ${\cal S}_{DT}'$ does not depend on the parameter $\tau$. The variation with $l/a$ of the universal part ${\cal S}_{UT}$ and ${\cal S}_{UT}'$ is shown by Fig.\ref{F9}. 

We would like to focus on the behavior in the curves of the entanglement entropy functional ${\cal S}_{UT}$ and  ${\cal S}_{UT}'$ near $au_{\ast} \sim \tau\;(u_{\ast} \sim u_{T})$. As shown in Fig.\ref{F9}, the value of the entanglement entropy functional ${\cal S}_{UT}'$ also increases with the increase of $l/a$. As is clear from the fact that the slope of the curve of ${\cal S}_{UT}'$ is almost flat, however, the increase rate of ${\cal S}_{UT}'$ with respect to $l/a$ is smaller than that of ${\cal S}_{UT}$. Therefore, the curves of the entanglement entropy ${\cal S}_{UT}$ and  ${\cal S}_{UT}'$ cross at $l=l_{crit}>l_{min}$. This fact exhibits that the entanglement entropy is governed by the configuration of the piecewise smooth surface for $l>l_{crit}$, since the Ryu-Takayanagi prescription requires the selection of curved surfaces with a smaller area. In other words, there is a transition for the entanglement entropy in the noncommutative theory. It has been shown that such transitions do not occur in corresponding (four-dimensional) commutative theory \cite{IBAFLAPZCATE}.

\begin{figure}[H]
\centering
\hspace*{0mm}
\includegraphics[width=80mm]{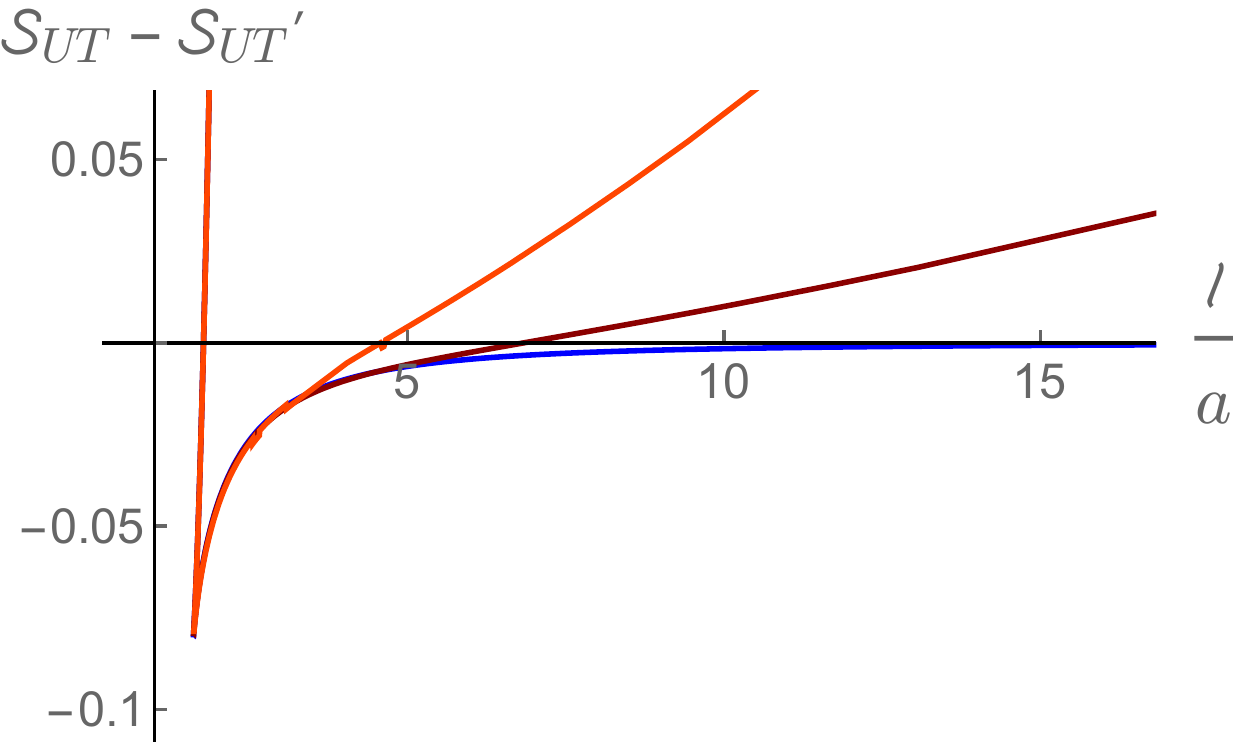} 
\caption{The variation with $l/a$ of the difference ${\cal S}_{UT}-{\cal S}_{UT}'$. The blue, brown and orange curve lines correspond respectively to $\tau=0.00,\;0.20$ and $0.30$.}
\label{F10}
\end{figure}

The variation with $l/a$ of the difference ${\cal S}_{UT}-{\cal S}_{UT}'$ is shown by Fig.\ref{F10}. 

${\cal S}_{UT}-{\cal S}_{UT}'=0$ means the cross of curves of the entanglement entropy ${\cal S}_{UT}$ and ${\cal S}_{UT}'$. Fig.\ref{F10} shows that the transition for the entanglement entropy occurs at positive $\tau$, whereas the transition for the entanglement entropy does not occur at $\tau=0$. The configuration of the smooth surface has the lowest entanglement entropy for small $\tau$ and is the dominant contribution. On the other hand, the configuration of the piecewise smooth surface has the lowest entanglement entropy for large $\tau$ and becomes the dominant contribution. 

%
%
\section{Conclusions and discussions}
\setcounter{section}{4}
\setcounter{equation}{0}
\addtocounter{enumi}{1}

In this paper, we have examined the properties of the holographic entanglement entropy in the holographic dual of the noncommutative Yang-Mills theory. The finite part of the holographic entanglement entropy in the noncommutative Yang-Mills theory can be derived without cutoff dependence, and thus is universal. Although the divergence part of the holographic entanglement entropy in the commutative regime ($au_{\ast} \to 0$ limit) follows the area law, that in the deep noncommutative regime ($au_{\ast} \to \infty$ limit) follows the volume law. This {\it area/volume} law transition \cite{LBCF} could be understood as a feature of nonlocal field theories. 

The universal part of the holographic entanglement entropy as a function of length $l$ in the noncommutative theory exhibits a peculiar behavior. There exists a minimum length $l_{min}$ in noncommutative theory, and the curve of the entanglement entropy branches at points of the minimum length $l_{min}$. This behavior seems to be a remarkable feature that somehow reflects the {\it area/volume} law transition.

The holographic entanglement entropy in the deep noncommutative regime does not satisfies the strong subadditivity inequality. In addition, the value of the mutual information of any two regions of space in the deep noncommutative regime takes the negative value for all values of the distance. These undesired results might be related to the nonlocal properties of the noncommutative Yang-Mills theory. It needs further discussion on these points. It should be emphasized that the noncommutativity of space itself does not violate the strong subadditivity. The value of the mutual information in the noncommutative theory under the condition of $au_{\ast}<0.7946$ takes a rather large value compared to that in the commutative regime.

The entropic $c$-function in the noncommutative theory does not satisfy monotonicity with respect to  $l$ as the scale parameter. The monotonicities of the $c$-functions build from the entanglement entropy are derived as a result of the strong subadditivity and Lorentz invariance of the theory. This property of the entropic $c$-function in the noncommutative theory might be understood from the breaking Lorentz symmetry. On the other hand, the behavior of the entropic c-function in the IR limit of the commutative regime is similar to that in the UV limit of the deep noncommutative regime, if we interpret $l \to \infty$ as infrared limit and $l \to 0$ as ultraviolet limit. This phenomenon seems to be a kind of the UV/IR relation. It would be interesting to discuss such arguments with the entropic $c$-theorems in four dimensions ($a$-theorem) \cite{SNS, HCETGT}.

There exist a minimum length $l_{min}$ that gives a branch point of the curve of the entanglement entropy even in the noncommutative theory at finite temperature. On the other hand, the effect of temperature on the behavior of the entanglement entropy curve becomes remarkable for large $l/a\;\;(u_{\ast} \sim u_{T})$. 
It is also notice that there is a transition from the configuration of the smooth surface to the piecewise smooth surface for the entanglement entropy in the noncommutative theory. The anisotropy of the noncommutativity is considered to be a major factor in inducing this transition.

The holographic entanglement entropy can act as an order parameter for the confinement/deconfinement phase in a confining gauge theory and can be used as a diagnostic tool to examine the confinement/deconfinement phase structure 
\footnotemark[3]\mbox{${}^{)}$} \cite{TNTT, IRKDKAM}. It is interesting to discuss how the noncommutativity parameter influences the confinement/deconfinement phase structure. 
In addition, it has been known that the existence of ``stripe phase", that is characterized by the peculiar position-dependence of the order parameter in the noncommutative scalar field theories \cite{SSGSLS, PCDZ}. It seem that such phases do not yet manifest in the holographic entanglement entropy of the noncommutative Yang-Mills theory. We hope to discuss these subject in the future. 

\footnotetext[3]{${}^{)}$ It was also pointed out that the entanglement measures may be unable to discriminate confining theories from non-confining ones with a mass gap \cite{NJJGS}.} 
%


\clearpage 

%
%

\end{document}